\documentclass[12pt]{article}
\usepackage{amsmath}
\usepackage{amsfonts}
\usepackage{amssymb,amsthm}
\usepackage{hyperref}
\usepackage{authblk}
\usepackage{graphicx}
\usepackage{subcaption}
\usepackage[english]{babel}
\usepackage[usenames]{color}
\usepackage{xcolor}
\usepackage{booktabs}
\usepackage{geometry}

\title{Testing the Ricci Focusing Degeneracy between Dynamical Dark Energy and Matter Inhomogeneity}
\date{\vspace{-5ex}}

\author{
     I.A.~Moiseev$^{1}$\thanks{E-mail: lxyniti@gmail.com}, O.S.~Sazhina$^{1}$\thanks{E-mail: cosmologia@yandex.ru}
 }

 \date{
     {\small
     $^{1}$Sternberg Astronomical Institute, Moscow, Russia\\[1ex]}
    \vspace{0.5em} 
     \today
 }

\begin{document}

\maketitle

\begin{abstract}

The influence of dynamical dark energy on light propagation within the framework of the Zeldovich--Kantowski--Dyer--Roeder (ZKDR) approximation is discussed. This effect is considered in terms of the redshift-dependent parameter $\alpha(z)$ of the ZKDR model. 
The total matter density is assumed to consist of two parts: one that is homogeneous, accounting for a fraction $\alpha(z)$  of the total density, and a second, inhomogeneous component, accounting for the remaining fraction $(1-\alpha(z))$.
We show, using an analytical study of the low-redshift asymptotics together with independent parameter fits to DESI BAO and Pantheon+ data, that dynamical dark energy with $w_0 \neq -1$ (in the CPL parametrization) at low redshift cannot be mimicked by the regular ZKDR smoothness parameterizations considered. For ordinary ZKDR on a flat $\Lambda$CDM background with the same $H_0$ and $\Omega_m$ as the CPL model, the reconstructed effective parameter $\alpha_{\rm eff}(z)$ exhibits a singular low-redshift behavior, whereas the phenomenological ZKDR forms remain finite. For the modified ZKDR prescription, the corresponding effective parameter has a finite low-redshift limit, but it falls out of a physically meaningful range for the adopted CPL parameters.

\end{abstract}
\section{Introduction}
\noindent

The inhomogeneous distribution of matter in space affects a number of cosmological problems related to light propagation. These problems include: (1) the discrepancy between the present-day values of the Hubble parameter based on observational data from the ``near Universe'' and from the early Universe (Hubble tension problem); (2) the calculation of angular diameter distances to remote sources. 

The recent results of the DESI survey, suggesting that dark energy may be dynamical with $w<-1$ \cite{DESI, Ormondroyd, Berti}, pose a new challenge for modern cosmology, whose resolution may require revising the Standard cosmological model ($\Lambda$CDM). The existence of dark energy in the form of phantom energy in the Universe opens up the possibility of stable exotic structures, such as wormholes \cite{zhetp}, but more importantly, raises the question of whether our current understanding of the Universe remains self-consistent in the presence of violations of the energy conditions. Thus, the problem of light propagation in the Universe filled with dynamical dark energy with a nonzero $p_\text{DE}+c^2\rho_\text{DE}$ term requires a detailed analysis.

Observations of sources at cosmological distances are strongly affected by inhomogeneities of the Universe.
Since inhomogeneities of matter (baryonic and dark) and dark energy can affect the trajectories of light rays, the task of distinguishing between these contributions becomes important, as does the question of whether the matter distribution can mimic a form of dynamical dark energy, and whether it is possible to dispense with the concept of dark energy altogether. So, the parameter which models the degree of lumpiness \cite{LTB} could, in principle, be statistically consistent with DESI observational data, similarly to the predictions of a cosmological model with dynamical dark energy based on DESI and Planck CMB data, except at low redshifts.

The case of inhomogeneous pressureless matter and dark energy with a constant equation-of-state parameter was discussed in \cite{Linder 1988}--\cite{Sereno 2001}. The authors consider quintessence with a constant $w \neq -1$, i.e., dark energy with a varying density. It was shown in \cite{Sereno 2002} that for a given redshift and matter density parameter $\Omega_m$ the cosmological constant in the $\Lambda$CDM  model could be interpreted as quintessence only due to the gravitational lensing by large-scale structure. The authors consider quintessence with $w \neq -1$ and show that, due to lensing dispersion, at $\Omega_m = 0.4$, the cosmological constant at $z = 1$ is indistinguishable from quintessence with $w < -0.84$ (at $2\sigma$ level).


However, questions about small redshifts and dynamical dark energy remained open.  

In our paper, we show that for $w_0\neq-1$ and the fixed background parameters, the dynamical dark energy distance near $z=0$ cannot be mimicked by the regular ZKDR smoothness parameterizations considered.

The paper is organized as follows. In Section 2, we describe the model and assumptions. In Section 3, we present a theoretical comparison of the ZKDR and dynamical dark energy models and derive the effective parameter \(\alpha_{\rm eff}\). In this section, we also use an observational test with DESI BAO and Pantheon+ to see whether the low-redshift difference seen in the theoretical comparison survives
when the model parameters are inferred directly from observations. We discuss the low-redshift asymptotics. Section 4 considers the modified ZKDR equation. Sections 5 and 6 present the discussion and conclusions.

\section{The models and assumptions}
\label{sec:bm}
\noindent

Here we specify the model and the numerical values of the cosmological parameters, as well as give the basic equations used in the paper.

Within the $\Lambda$CDM model the conventional approach for treating the problem of light propagation is based on solving the Sachs optical equations within the Friedmann--Robertson--Walker (FRW) homogeneous and isotropic background.
This approach describes the dependence of the angular diameter distance $D_A(z)$ on redshift $z$. Dark energy is represented by the cosmological constant $\Lambda$ or, equivalently, a field with the equation of state $p_\text{DE}=-c^2\rho_\text{DE}$. 

Light propagation is affected not only by the background expansion of the Universe but also by local matter inhomogeneities. Deviations from a homogeneous matter distribution are expected to affect light propagation through distortions of bundles of light geodesics. This phenomenon is known as the Zeldovich effect and in the modern literature is commonly referred to as the ZKDR approximation (Zeldovich--Kantowski--Dyer--Roeder). The Zeldovich effect for the Universe filled with dark matter and dark energy has been studied in a number of papers (see  \cite{1}--\cite{Bolejko} and references therein).

The ZKDR equation for the angular diameter distance $D_A(z)$ is given by \cite{ZKDR}:
\begin{equation}
\label{Eq:ZKDR}
\frac{d^2D_A(z)}{dz^2}+\left(\frac{d \ln H(z)}{dz}+\frac{2}{1+z}\right)\frac{dD_A(z)}{dz}=-\frac{3}{2}\Omega_m\frac{H^2_0}{H(z)^2}(1+z)\alpha(z)D_A(z),    
\end{equation}
where the total matter density is assumed to consist of two parts: one that is homogeneous, accounting for a fraction $\alpha(z)$ of the total density, and the second ($1-\alpha(z)$), which consists of matter inhomogeneities.

The corresponding Hubble parameter as a function of redshift is expressed as follows:
\[
H(z)=H_{\Lambda\text{CDM}} = H_0 \sqrt{\Omega_m (1+z)^3 + \Omega_\Lambda}.
\]

Here, $\Omega_\Lambda$ and $\Omega_m$ are the dark energy and matter (baryonic and dark) parameter, respectively. We neglect the curvature term. If $\alpha(z)=1$ then the model reduces to the FRW distance-redshift relation. 
    
For our purpose of defining an \(\alpha(z)\) associated with dynamical dark energy, we derive a similar equation from the Sachs optical equations \cite{Sachs} specifying the energy-momentum tensor:
\[
\frac{d^2D_A}{ds^2}=-\left(|\sigma|^2+\frac{1}{2}R_{\alpha\beta}k^\alpha k^\beta\right)D_A.
\]

For the Ricci tensor $R_{\alpha\beta}$ from Einstein's equations one obtains:
\[
R_{\alpha\beta}k^\alpha k^\beta-\frac{1}{2}Rg_{\alpha\beta}k^\alpha k^\beta=8\pi GT_{\alpha\beta}k^\alpha k^\beta,
\]
where
\[
g_{\alpha\beta}k^\alpha k^\beta=0
\]
since $k^\alpha k_\alpha=g_{\alpha\beta}k^\alpha k^\beta=0$. (Hereinafter we consider $c=1$ for clarity.)

To describe the dynamical dark energy the energy-momentum tensor is assumed to be that of a perfect fluid:
\[
T_{\alpha\beta}=(p+\rho)u^\alpha u^\beta+pg_{\alpha\beta}.
\]

Then one finds
\[
R_{\alpha\beta}k^\alpha k^\beta=8\pi G(p+\rho)(u_\alpha k^\alpha)^2,
\]
and for the baryonic and dark matter (denoted in total as $\rho_m$), as well as for the dark energy ($\rho_{\text{DE}}$) we can write
\[
p+\rho=\rho_m+\rho_{\text{DE}}+p_{\text{DE}}=\rho_m+\rho_{\text{DE}}(1+w).
\]

The equation for the angular diameter distance $D_A$ becomes as follows:
\[
\frac{d^2D_A}{ds^2}=-|\sigma|^2D_A-4\pi G[\rho_m+\rho_{\text{DE}}(1+w)](u_\alpha k^\alpha)^2D_A.
\]

Finally, let us switch from the affine parameter $s$ to the redshift $z$. Omitting the technical details from \cite{Sachs}, we immediately write down
\[
\frac{dz}{ds}=(1+z)^2H(z).
\]

Then the derivatives over $s$ become the derivatives with respect to $z$:
\[
\frac{d}{ds}=\frac{dz}{ds}\frac{d}{dz}=H(z)(1+z)^2\frac{d}{dz},
\]
\[
\frac{d^2}{ds^2}=H^2(z)(1+z)^4\left[\frac{d^2}{dz^2}+\left(\frac{1}{H(z)}\frac{dH(z)}{dz}+\frac{2}{1+z}\right)\frac{d}{dz}\right],
\]
and the resulting equation takes the following form:
\[
\frac{d^2D_A(z)}{dz^2}+\left(\frac{d \ln H(z)}{dz}+\frac{2}{1+z}\right)\frac{dD_A(z)}{dz}=
\]
\begin{equation}
\label{Eq:S_DESI}
    =-\frac{|\sigma|^2}{H(z)^2(1+z)^4}D_A(z)-\frac{4\pi G}{H(z)^2(1+z)^2}\left[ \rho_m+\rho_{\text{DE}}(1+w(z)) \right]D_A(z).
\end{equation}

When restoring the dimensional coefficients in conjunction with the initial conditions
\[D_A(z=0)=0, \frac{dD_A}{dz}(z=0)=\frac{c}{H_0}
\]
the equation can be solved.

Considering the dynamical dark energy model, for the equation of state parameter $w(z)$ we will use the standard CPL parametrization\cite{CPL}:
\begin{equation}
\label{Eq:st}
w=w(z)=w_0 + w_a \cdot \bigg(1- \dfrac{1}{1+z}\bigg),
\end{equation}
which for the dark energy with a general equation of state $p_\text{DE}=w\rho_\text{DE}$ yields the following evolution law \cite{L2003}:
\begin{align}
\label{Eq:rho_ph}
\rho_{\text{DE}}(z) &= \Omega_\text{DE} \cdot \dfrac{3H_0^2 c^2}{8\pi G} \cdot \exp \bigg( 3 \int_{0}^{z} \dfrac{1+w(z)}{1+z}dz \bigg) \\
&=\Omega_\text{DE} \cdot \dfrac{3H_0^2 c^2}{8\pi G} (1+z)^{3(1+w_0+w_a)} \exp \bigg( -\dfrac{3w_a z}{1+z} \bigg),\nonumber
\end{align}
where $\Omega_\text{DE} = 1 - \Omega_m = 0.6781 \pm 0.0065$, $H_0=66.71 \pm 0.62$km s$^{-1}$ Mpc$^{-1}$, the parameter values were adopted as $w_0=-0.804^{+0.063}_{-0.063}$, $w_a=-0.67^{+0.28}_{-0.23}$ (CMB+SDSS+DESY5+CC, \cite{w0wa} with all quoted uncertainties corresponding to 68\% CL and $w_0=-0.8$ being used in the numerical illustrations). These values are used for the numerical illustration in Section 4; the fits in Section 3.3 use independently sampled parameters. The speed of light $c$ here is restored explicitly in order to use this expression in the numerical calculations. 

Neglecting optical shear, we obtain the equation for the angular diameter distance in a dynamical dark energy background, that takes the form
\begin{equation}
\label{Eq:dDE}
\frac{d^2 D_A}{dz^2} + \left( \frac{d \ln H_{\text{DE}}}{dz} + \frac{2}{1+z} \right) \frac{d D_A}{dz} = -\frac{4\pi G}{H_{\text{DE}}^2 (1+z)^2} \left[ \rho_m + \rho_{\text{DE}} (1+w) \right] D_A,
\end{equation}
where the corresponding Hubble parameter is expressed as follows:
\[
H_{\text{DE}} = H_0 \sqrt{\Omega_m (1+z)^3 + \Omega_{\text{DE}} \exp \left( 3 \int_0^z \frac{1+w(z')}{1+z'} dz' \right)}.
\]


\section{Comparison with ZKDR}
\subsection{Effective parameter $\alpha_\text{eff}$}
\noindent

First, we examine whether a matter-inhomogeneity parameter \(\alpha(z)\) can mimic the effect of dynamical dark energy or, vice versa, whether the contribution of dynamical dark energy to Ricci focusing can be represented in terms of an effective ZKDR inhomogeneity parameter. To compare these effects, we rewrite the corresponding equations for the angular diameter distance $D_A$ in the following way.

The ordinary ZKDR equation for a flat $\Lambda$CDM background is
\begin{equation}
\nonumber
\frac{d^2 D_A}{dz^2} + \left( \frac{d \ln H_{\Lambda\text{CDM}}(z)}{dz} + \frac{2}{1+z} \right) \frac{d D_A}{dz} = -\frac{3}{2} \Omega_m \frac{H_0^2}{H_{\Lambda\text{CDM}}^2(z)} (1+z) \alpha(z) D_A,
\end{equation}

The corresponding equation for a dynamical dark energy:
\begin{equation}
\nonumber
\frac{d^2 D_A}{dz^2} + \left( \frac{d \ln H_{\text{DE}}}{dz} + \frac{2}{1+z} \right) \frac{d D_A}{dz} = -\frac{4\pi G}{H_{\text{DE}}^2 (1+z)^2} \left[ \rho_m + \rho_{\text{DE}} (1+w(z)) \right] D_A,
\end{equation}
where the corresponding Hubble parameters are expressed as follows:
\begin{align*}
H_{\Lambda\text{CDM}} &= H_0 \sqrt{\Omega_m (1+z)^3 + \Omega_\Lambda}, \\[0.5em]
H_{\text{DE}} &= H_0 \sqrt{\Omega_m (1+z)^3 + \Omega_{\text{DE}} \exp \left( 3 \int_0^z \frac{1+w(z')}{1+z'} dz' \right)}.
\end{align*}

To find a matter-inhomogeneity parameter \(\alpha_{\rm eff}(z)\) that, in a flat background, mimics the presence of dynamical dark energy, one should consider the following equation:

\begin{equation*}
\frac{d^2 D_A^{\text{DE}}}{dz^2} + \left( \frac{d \ln H_{\Lambda\text{CDM}}}{dz} + \frac{2}{1+z} \right) \frac{d D_A^{\text{DE}}}{dz} = -\frac{3}{2} \Omega_m \frac{H_0^2}{H_{\Lambda\text{CDM}}^2} (1+z) \alpha_{\text{eff}}(z) D_A^{\text{DE}},
\end{equation*}
from which one obtains, considering the distances $D_A^\text{DE}$ calculated from the equation (\ref{Eq:dDE}), the following expression for the effective inhomogeneity parameter
\begin{equation}
\label{Eq:alpha_eff}
\alpha_{\text{eff}}(z) = -\frac{2 H_{\Lambda\text{CDM}}^2(z)}{3 \Omega_m H_0^2 (1+z) D_A^{\text{DE}}} \left[ \frac{d^2 D_A^{\text{DE}}}{dz^2} + \left( \frac{d \ln H_{\Lambda\text{CDM}}}{dz} + \frac{2}{1+z} \right) \frac{d D_A^{\text{DE}}}{dz} \right].
\end{equation}

This expression defines an effective ZKDR parameter $\alpha_{\rm eff}(z)$ such that, when inserted into the ZKDR equation in a flat $\Lambda$CDM background with the same $H_0$ and $\Omega_m$ as the dynamical-dark-energy model, it reproduces the angular diameter distance $D_A^{\rm DE}(z)$ generated by that model.

Having established this formal framework, we have not yet determined whether this effective parameter \(\alpha_{\rm eff}(z)\) can be represented by the phenomenological matter-inhomogeneity parameterizations used in ZKDR models.

We compare the reconstructed function with three phenomenological smoothness parameterizations. The first two functional forms are discussed in \cite{ZKDR}, while the third is taken from \cite{Bolejko}. In this section, all three are used in the ordinary ZKDR equation:

\begin{itemize}
    \item ZKDR1:
    \begin{equation}
    \alpha(z)=\alpha_0+\alpha_1 z;
    \end{equation}

    \item ZKDR2:
    \begin{equation}
    \alpha(z) = \frac{\beta_0(1+z)^{3\gamma}}
    {1+\beta_0(1+z)^{3\gamma}};
    \end{equation}

    \item ZKDR3 \cite{Bolejko}:
    \begin{equation}
    \alpha(z) = 1+\frac{\delta}{(1+z)^{5/4}}.
    \end{equation}
\end{itemize}

Equation (\ref{Eq:alpha_eff}) defines the smoothness function required to reproduce the specified dynamical-dark-energy distance in the ordinary ZKDR equation. We next examine whether this function remains regular and belongs to the usual smoothness interval near \(z=0\).

\subsection{Low-redshift asymptotics}
\noindent

Let us study analytically the asymptotic behavior of the effective parameter $\left. \alpha_{\text{eff}}(z) \right|_{z \to 0}$.

Expanding
the angular diameter distance around $z=0$ gives
\begin{align*}
D_A(z) &= D_A(0) + D_A'(0) z + \frac{1}{2} D_A''(0) z^2 + \mathcal{O}(z^3) \\
D_A(0) &= 0 \\
D_A'(0) &= \frac{c}{H_0}
\end{align*}

At $z=0$, Eq.~(\ref{Eq:dDE}) gives
\begin{equation*}
D_A''(0) = -\left( \frac{d \ln H_{\text{DE}}}{dz} + 2 \right) D_A'(0) = -\left( \left. \frac{d \ln H_{\text{DE}}}{dz} \right|_{z=0} + 2 \right) \frac{c}{H_0}
\end{equation*}


For the dynamical dark energy Hubble parameter,

\[
\frac{d\ln H_{\rm DE}}{dz}
=
\frac{
3\Omega_m(1+z)^2
+
3\Omega_{\rm DE}
\dfrac{1+w(z)}{1+z}
\exp\left[
3\int_0^z\dfrac{1+w(z')}{1+z'}\,dz'
\right]
}{
2\left[
\Omega_m(1+z)^3
+
\Omega_{\rm DE}
\exp\left[
3\int_0^z\dfrac{1+w(z')}{1+z'}\,dz'
\right]
\right]
},
\]
so that

\[
\left.\frac{d\ln H_{\rm DE}}{dz}\right|_{z=0}
=
\frac{3\Omega_m+3\Omega_{\rm DE}(1+w_0)}
{2(\Omega_m+\Omega_{\rm DE})}.
\]

For $\Lambda$CDM,

\[
\frac{d\ln H_{\Lambda{\rm CDM}}}{dz}
=
\frac{3\Omega_m(1+z)^2}
{2[\Omega_m(1+z)^3+\Omega_\Lambda]},
\]
and therefore

\[
\left.\frac{d\ln H_{\Lambda{\rm CDM}}}{dz}\right|_{z=0}
=
\frac{3\Omega_m}{2(\Omega_m+\Omega_\Lambda)}.
\]

Assuming the same $H_0$ and $\Omega_m$ in both flat models, so that $\Omega_\Lambda=\Omega_{\rm DE}=1-\Omega_m$, one finds

\[
D_A^{{\rm DE}\,''}(0)
+
\left[
\left.\frac{d\ln H_{\Lambda{\rm CDM}}}{dz}\right|_{z=0}+2
\right]
D_A^{{\rm DE}\,'}(0)
=
-\frac{3c}{2H_0}
\frac{\Omega_{\rm DE}(1+w_0)}
{\Omega_m+\Omega_{\rm DE}}.
\]

Since
\[
D_A(z)\simeq\frac{cz}{H_0},
\qquad z\rightarrow0,
\]

Eq.(\ref{Eq:alpha_eff}) yields
\begin{equation}
\label{Eq:alpha_lowz}
\alpha_{\rm eff}(z)
\simeq
\frac{\Omega_{\rm DE}}{\Omega_m}(1+w_0)\frac{1}{z},
\qquad z\rightarrow0.
\end{equation}

\[
\lim_{z \to 0} \alpha_{\text{eff}}(z) = \lim_{z \to 0} \frac{\Omega_{\text{DE}}}{\Omega_m} (1+w_0) \frac{1}{z} = +\infty\quad \text{for } w_0>-1
\]

By contrast, the phenomenological ZKDR parameterizations have finite low-redshift limits,
\begin{align}
\alpha_{\rm ZKDR1}(0)&=\alpha_0,
\nonumber\\
\alpha_{\rm ZKDR2}(0)&=\frac{\beta_0}{1+\beta_0},
\nonumber\\
\alpha_{\rm ZKDR3}(0)&=1+\delta.\nonumber
\end{align}

\subsection{Observational test with DESI BAO and Pantheon+}
\label{sec:mcmc}
\noindent

We next fit the CPL model and the three ordinary ZKDR models
independently to DESI DR1 BAO and Pantheon+ distance data.
Unlike the analytical comparison above, these fits allow each model to have its own cosmological parameters. The resulting posterior distributions are then used to compare the reconstructed CPL smoothness function with the phenomenological ZKDR functions. For the BAO likelihood, we used the DESI DR1 transverse-distance measurements $D_M/r_d$ listed in Table \ref{tab:bao_data}.

\begin{table}[ht]
\centering
\caption{DESI DR1 transverse-distance measurements
with $1\sigma$ uncertainties \cite{DESI_DR1}}
\label{tab:bao_data}
\begin{tabular}{cc}
\hline
$z$ & $D_M/r_d$ \\
\hline
0.510 & $13.62 \pm 0.25$ \\
0.706 & $16.85 \pm 0.32$ \\
0.930 & $21.71 \pm 0.28$ \\
1.317 & $27.79 \pm 0.69$ \\
2.330 & $39.71 \pm 0.94$ \\
\hline
\end{tabular}
\end{table}

The Pantheon+ sample was used to constrain the luminosity-distance--redshift relation. The sound horizon at the drag epoch $r_d$ was treated as a free parameter in all fits.

For the CPL dynamical-dark-energy model, the sampled cosmological parameters were

\begin{equation}
(\Omega_m,\,h,\,w_0,\,w_a,\,r_d)
\end{equation}
for the BAO-only analysis and

\begin{equation}
(\Omega_m,\,h,\,w_0,\,w_a,\,r_d,\,M_B)
\end{equation}
for the joint DESI BAO + Pantheon+ analysis. For the ZKDR models, the background expansion was taken to be $\Lambda$CDM, while the parameters of $\alpha(z)$ were sampled simultaneously with $\Omega_m$, $h$, $r_d$, and $M_B$. It is worth noting that the data do not constrain $h$, $r_d$, and $M_B$ independently: the BAO measurements of $D_M/r_d$ constrain the combination $h r_d$, while the Pantheon+ sample constrains $M_B - 5\log_{10}h$. Consequently, the marginal posterior estimates of these parameters depend on the adopted priors and should not be interpreted as separate measurements of $h$, $r_d$, and $M_B$.

Uniform priors used in the analysis are listed in Table \ref{tab:priors}. These priors do not enforce \(0\leq\alpha(z)\leq1\) for all parameterizations and therefore the fits include phenomenological extensions beyond the usual smoothness interval.

\begin{table}[ht]
\centering
\caption{Priors}
\label{tab:priors}
\begin{tabular}{lll}
\toprule
Parameter & Model & Range \\
\midrule
$\Omega_m$ & all & $(0.1,\,0.5)$ \\
$h$ & all & $(0.5,\,0.8)$ \\
$w_0$ & CPL & $(-2.5,\,0.0)$ \\
$w_a$ & CPL & $(-3.0,\,2.0)$ \\
$r_d$ [Mpc] & all & $(130,\,160)$ \\
$M_B$ & all joint fits & $(-20.5,\,-18.0)$ \\
$\alpha_0$ & ZKDR1 & $(0.0,\,1.0)$ \\
$\alpha_1$ & ZKDR1 & $(0.0,\,1.0)$ \\
$\beta_0$ & ZKDR2 & $(0.0,\,1.0)$ \\
$\gamma$ & ZKDR2 & $(0.0,\,1.0)$ \\
$\delta$ & ZKDR3 & $(-2.0,\,2.0)$ \\
\bottomrule
\end{tabular}
\end{table}

The sampled posterior distributions are summarized in
Table \ref{tab:posterior_summary}. For Pantheon+, the full statistical and systematic covariance matrix was used, while the DESI transverse-distance likelihood was constructed from the quoted \(D_M/r_d\) uncertainties.

\begin{table}[h!]
\centering
\caption{Posterior parameter estimates (median and 68\% credible interval) for the CPL dark-energy model and the three ZKDR phenomenological focusing models jointly fit to DESI DR1 BAO\cite{DESI_DR1} and Pantheon+\cite{Pantheon+1, Pantheon+2}.}
\label{tab:posterior_summary}
\begin{tabular}{l l c}
\hline\hline
Model & Parameter & Value \\
\hline
CPL (DESI BAO only) & $\Omega_m$ & $0.2941^{+0.0716}_{-0.0889}$ \\
 & $h$ & $0.6763^{+0.0662}_{-0.0628}$ \\
 & $w_0$ & $-0.6168^{+0.4148}_{-0.5738}$ \\
 & $w_a$ & $-1.0053^{+1.3457}_{-1.1963}$ \\
 & $r_d$ & $144.5985^{+10.0532}_{-9.8170}$ \\
\hline
CPL (DESI BAO + Pantheon+) & $\Omega_m$ & $0.2697^{+0.0766}_{-0.0916}$ \\
 & $h$ & $0.6924^{+0.0498}_{-0.0477}$ \\
 & $w_0$ & $-0.8486^{+0.0899}_{-0.1135}$ \\
 & $w_a$ & $0.0756^{+0.5508}_{-0.9953}$ \\
 & $r_d$ & $143.9204^{+10.4664}_{-9.7142}$ \\
 & $M_B$ & $-19.3710^{+0.1517}_{-0.1539}$ \\
\hline
ZKDR1 (Joint DESI+Pantheon+) & $\Omega_m$ & $0.3754^{+0.0307}_{-0.0288}$ \\
 & $h$ & $0.6906^{+0.0517}_{-0.0470}$ \\
 & $\alpha_0$ & $0.4760^{+0.2414}_{-0.2788}$ \\
 & $\alpha_1$ & $0.2832^{+0.2430}_{-0.2002}$ \\
 & $r_d$ & $144.5730^{+10.6181}_{-10.0602}$ \\
 & $M_B$ & $-19.3741^{+0.1561}_{-0.1544}$ \\
\hline
ZKDR2 (Joint DESI+Pantheon+) & $\Omega_m$ & $0.3802^{+0.0248}_{-0.0231}$ \\
 & $h$ & $0.6902^{+0.0503}_{-0.0468}$ \\
 & $\beta_0$ & $0.7414^{+0.1847}_{-0.2565}$ \\
 & $\gamma$ & $0.7313^{+0.1889}_{-0.2594}$ \\
 & $r_d$ & $144.7464^{+10.5644}_{-10.0381}$ \\
 & $M_B$ & $-19.3755^{+0.1536}_{-0.1521}$ \\
\hline
ZKDR3 (Joint DESI+Pantheon+) & $\Omega_m$ & $0.3666^{+0.0339}_{-0.0324}$ \\
 & $h$ & $0.6924^{+0.0493}_{-0.0486}$ \\
 & $\delta$ & $-0.4855^{+0.3052}_{-0.2883}$ \\
 & $r_d$ & $144.0764^{+10.9130}_{-9.5250}$ \\
 & $M_B$ & $-19.3698^{+0.1491}_{-0.1584}$ \\
\hline\hline
\end{tabular}
\end{table}

\begin{figure}
    \centering
    \includegraphics[width=0.9\linewidth]{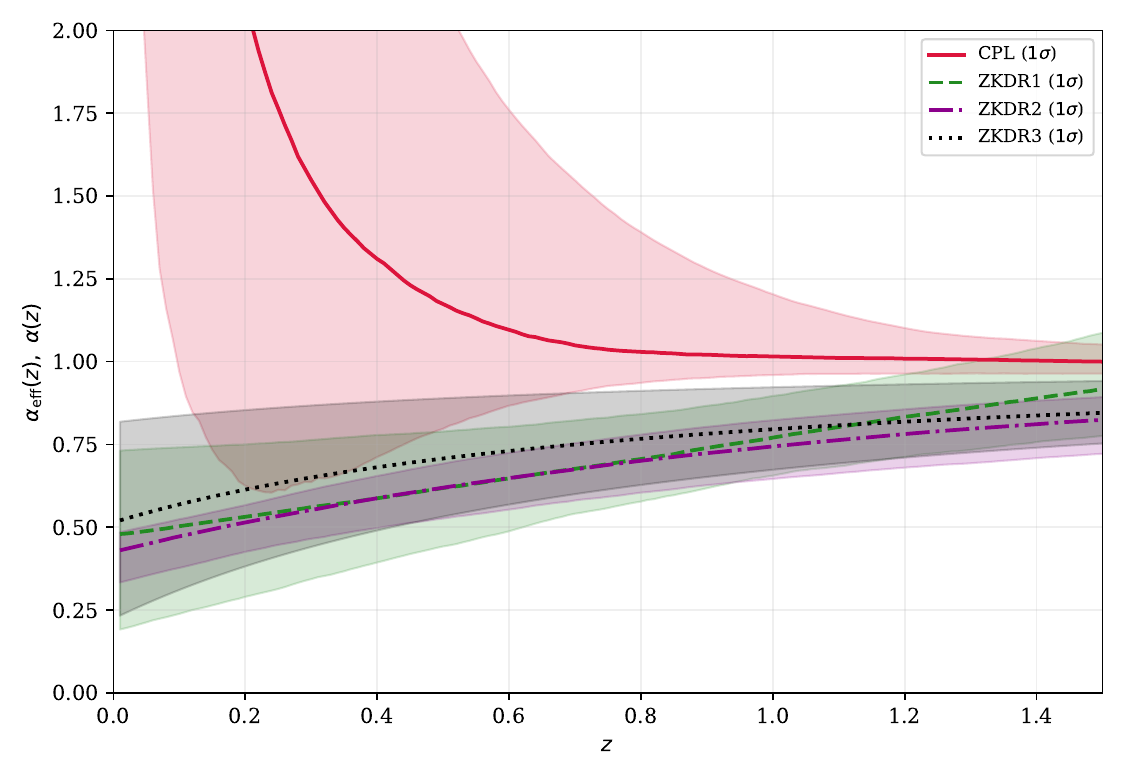}
    \caption{Observationally constrained comparison of the effective focusing parameter reconstructed from the joint CPL fit with the ZKDR1--ZKDR3 phenomenological parameterizations. The solid curves and shaded regions show the posterior median and \(1\sigma\) credible interval.}
    \label{fig:alpha_obs}
\end{figure}

The addition of Pantheon+ data substantially narrows the allowed range of the present-day CPL parameter $w_0$ compared with the DESI BAO-only fit, whereas $w_a$ remains considerably less constrained. For each sample of the joint CPL posterior, we reconstructed the corresponding effective parameter $\alpha_\text{eff}(z)$ using Eq.(\ref{Eq:alpha_eff}).
For the ZKDR models, the posterior samples were propagated through their respective $\alpha(z)$ parameterizations. This yields an observationally constrained comparison directly in the same effective matter-inhomogeneity description, presented in Fig.\ref{fig:alpha_obs}.

The observational reconstruction reproduces the same qualitative feature that appeared in the theoretical description. The models can overlap at some redshift ranges, but the dynamical dark energy effective reconstruction shows a distinct low-redshift evolution that is not followed by the ZKDR approximations.

We next examine how the effective reconstruction
changes when $\alpha(z)$ also enters the coefficients of the
modified ZKDR equation.

\section{Comparison with the modified ZKDR equation}
\noindent

The low-redshift behavior derived above follows from the ZKDR equation with a fixed $\Lambda$CDM expansion history $H(z)$. In this equation, $\alpha(z)$ enters only the Ricci focusing term. Let us now examine whether the same conclusion holds when the matter-inhomogeneity parameter also enters the expansion terms, as in the modified ZKDR equation considered in \cite{ZKDR}.

Omitting the curvature term, the modified ZKDR equation takes the form \cite{ZKDR}
\begin{align}
\label{Eq:mZKDR}
\frac{d^2D_A}{dz^2}
&+\left[
\frac{3\alpha(z)\Omega_m(1+z)^2}
{2[\alpha(z)\Omega_m(1+z)^3+\Omega_\Lambda]}
+\frac{2}{1+z}\right]\frac{dD_A}{dz}
\nonumber\\
&=-\frac{3\alpha(z)\Omega_m(1+z)}
{2[\alpha(z)\Omega_m(1+z)^3+\Omega_\Lambda]}D_A,
\end{align}
Here we follow the modified prescription of \cite{ZKDR}, which uses
\[
H_\alpha^2(z)=H_0^2\left[\alpha(z)\Omega_m(1+z)^3+\Omega_\Lambda\right]
\]
and neglects derivatives of $\alpha(z)$ in the distance equation.

To find the effective parameter which reproduces the same dynamical dark energy
distance, we substitute $D_A^{\rm DE}$ from Eq.~(\ref{Eq:dDE}) into
Eq.~(\ref{Eq:mZKDR}). Multiplying by
$\alpha(z)\Omega_m(1+z)^3+\Omega_\Lambda$ and collecting the terms with
$\alpha(z)$ gives
\begin{align*}
&\Omega_\Lambda\left[
\frac{d^2D_A^{\rm DE}}{dz^2}
+\frac{2}{1+z}\frac{dD_A^{\rm DE}}{dz}\right]
\\
&\quad+\alpha(z)\Omega_m\left[
(1+z)^3\frac{d^2D_A^{\rm DE}}{dz^2}
+\frac{7}{2}(1+z)^2\frac{dD_A^{\rm DE}}{dz}
+\frac{3}{2}(1+z)D_A^{\rm DE}\right]=0.
\end{align*}
The corresponding effective parameter is therefore
\begin{equation}
\label{Eq:alpha_eff_mZKDR}
\alpha_{\rm eff}^{\rm mZKDR}(z)=
-\frac{\displaystyle\Omega_\Lambda\left[
\frac{d^2D_A^{\rm DE}}{dz^2}
+\frac{2}{1+z}\frac{dD_A^{\rm DE}}{dz}\right]}
{\displaystyle\Omega_m\left[
(1+z)^3\frac{d^2D_A^{\rm DE}}{dz^2}
+\frac{7}{2}(1+z)^2\frac{dD_A^{\rm DE}}{dz}
+\frac{3}{2}(1+z)D_A^{\rm DE}\right]}.
\end{equation}
This expression and Eq. (\ref{Eq:alpha_eff}) reproduce the same distance using different propagation equations, therefore these effective parameters cannot be substituted for one another.

We first examine the low-redshift limit of Eq. (\ref{Eq:alpha_eff_mZKDR}). Using the distance expansion
from the preceding section, we obtain
\[
D_A^{\rm DE}(0)=0,\qquad
D_A^{{\rm DE}\,'}(0)=\frac{c}{H_0},\qquad
D_A^{{\rm DE}\,''}(0)=
-\left[2+\frac{3}{2}\bigl(\Omega_m+\Omega_{\rm DE}(1+w_0)\bigr)\right]
\frac{c}{H_0}.
\]
Substitution into Eq. (\ref{Eq:alpha_eff_mZKDR}) gives
\begin{equation}
\label{Eq:alpha_mZKDR_lowz}
\begin{aligned}
\lim_{z\to0^+}\alpha_{\rm eff}^{\rm mZKDR}(z)
&=
\frac{\Omega_\Lambda
[\Omega_m+\Omega_{\rm DE}(1+w_0)]}
{\Omega_m[1-\Omega_m-\Omega_{\rm DE}(1+w_0)]}=-\frac{1+w_0\Omega_{\rm DE}}{\Omega_m w_0}.
\end{aligned}
\end{equation}
For $0<\Omega_m<1$ and $w_0\ne0$, this limit is finite. Unlike the ordinary ZKDR equation, Eq.(\ref{Eq:mZKDR}) contains
$\alpha(z)$ in the coefficient multiplying the first derivative of the
distance. It can therefore change $D_A''(0)$ even when $\alpha(0)$ is
finite, removing the need for the $1/z$ term found for ordinary ZKDR.

However, a finite limit can lie outside the usual smoothness parameter interval $0\le\alpha\le1$. In particular,
\[
\lim_{z\to0}\alpha_{\rm eff}^{\rm mZKDR}(z)-1
=-\frac{1+w_0}{\Omega_m w_0},
\]
which is positive for $-1<w_0<0$. For the adopted parameters
$\Omega_m=0.3219$, $\Omega_{\rm DE}=0.6781$, $w_0=-0.80$ and
$w_a=-0.67$ from \cite{w0wa}, we obtain
\[
\lim_{z\to0}\alpha_{\rm eff}^{\rm mZKDR}(z)\simeq1.777.
\]
Thus, for this CPL distance and the specified background parameters, the modified equation replaces the $1/z$ behavior of the ordinary ZKDR reconstruction with a finite value outside the physically motivated smoothness parameter interval. 

\begin{figure}
    \centering
    \includegraphics[width=0.8\linewidth]{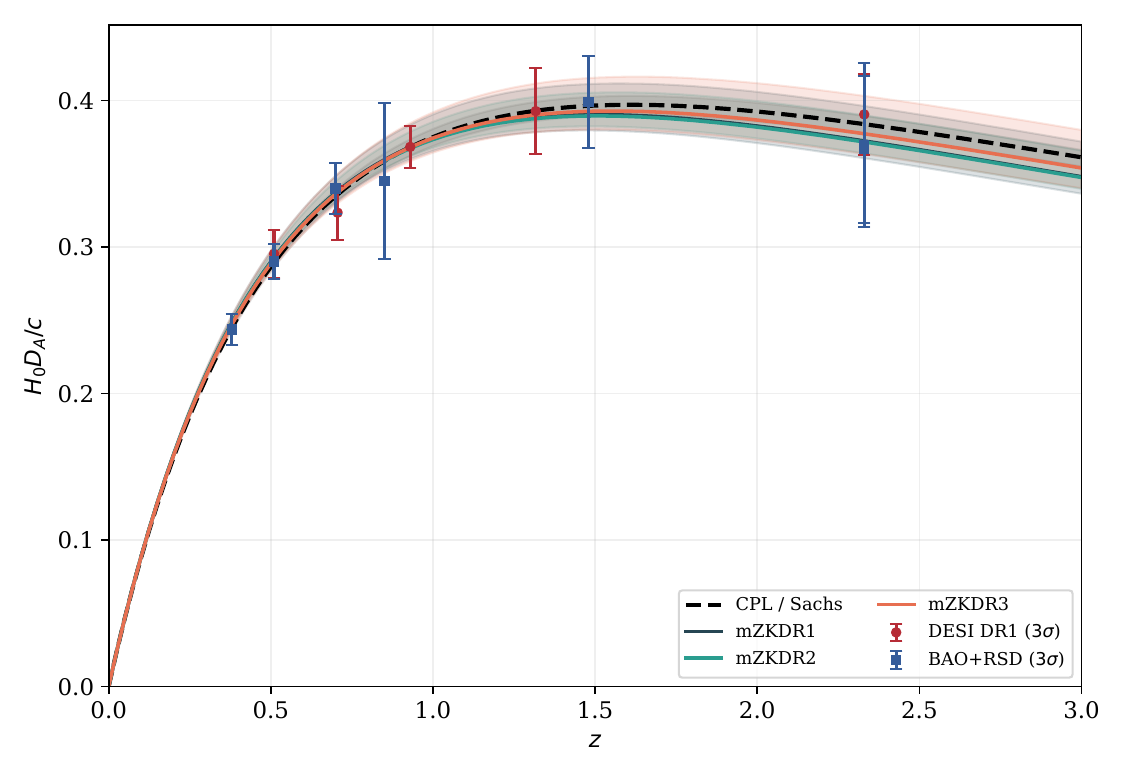}
    \caption{Normalized angular diameter distances $H_0D_A/c$ in the
    modified ZKDR models and in the CPL model. Each theoretical curve
    uses its own $H_0$. Shaded regions show $1\sigma$ intervals. DESI DR1 and BAO+RSD\cite{BAO} points are shown with $3\sigma$ uncertainties (with $r_d$ and $H_0$ fixed)}
    \label{fig:DA}
\end{figure}

For a numerical illustration, Fig.\ref{fig:DA} compares the normalized distances $H_0D_A/c$ obtained from Eq.(\ref{Eq:mZKDR}) with the CPL distance with CPL parameters taken from \cite{w0wa}. The mZKDR parameter estimates were taken from \cite{ZKDR}. The normalization uses the corresponding $H_0$ for each theoretical curve. The BAO points are converted using $D_A=(D_M/r_d)r_d/(1+z)$, with $r_d=147.16$~Mpc and $H_0=66.71$~km~s$^{-1}$~Mpc$^{-1}$ for their normalization. This normalization removes the overall distance scale from each theoretical curve. The BAO points are displayed at fixed $H_0r_d$ and are rather illustrative.

\section{Discussion}
\noindent
It was found in \cite{ZKDR} that the ZKDR and modified ZKDR models do not alleviate the Hubble tension when tested against Pantheon+ data with SH0ES calibration and H0LiCOW time-delay measurements. Our analysis addresses a different question: whether these approaches can reproduce the angular diameter distance of a dynamical dark energy model. We find a distinct $1/z$ behavior of the effective smoothness parameter for ordinary ZKDR models at low redshifts for $w_0\ne -1$ when the two models share the same fixed $H_0$ and $\Omega_m$, and for the modified prescription, the effective smoothness parameter has a finite low-redshift limit, but this limit lies outside the physically motivated region for the adopted CPL parameters. These results complement the observational findings of \cite{ZKDR} by identifying specific limitations of distance mimicry, instead of testing the resolution of the Hubble tension.

One should mention that the distinct low-redshift behavior obstructing the distance matching does not exclude approximate agreement over a finite redshift interval away from $z=0$. 

The joint BAO and supernova fits also involve an additional assumption. The ZKDR prescription describes light propagation along a narrow beam, whereas BAO distances are inferred from spatial clustering statistics. Applying the same \(\alpha(z)\) to both probes is therefore a phenomenological assumption whose validity depends on the relevant averaging procedure and we use these fits as an illustration within this prescription.

\section{Conclusion}
\noindent

In this paper we have studied the Ricci focusing effect in the presence of dynamical dark energy. The chosen framework of an effective clumpiness parameter $\alpha_{\rm eff}$ allowed us to show that although this phenomenological mapping can be established, it is not possible for a physically motivated ZKDR smoothness parameter to reproduce the specified dynamical dark energy distance near $z = 0$ under the adopted correspondence of background parameters.

The analytical asymptotic behavior of the effective ''clumping'' parameter explains the behavior found independently in both the theoretical and observational reconstructions. The discrepancy at low redshift is not caused by a particular choice of cosmological parameters. This discrepancy follows from the different low-redshift asymptotics and cannot be removed by adjusting the coefficients of a regular smoothness parameterization. Consequently, the formal expression given by Eq.\ref{Eq:alpha_eff} does not imply an observational equivalence between the dynamical-dark-energy model and the ZKDR1--ZKDR3 descriptions considered here.

The question of degeneracy at large redshifts $z>1$ remains open.

The independent fits provide a useful illustration: reconstructing $\alpha_{\rm eff}$ from the fitted CPL parameters under the same background correspondence preserves this low-redshift behavior.

Under the adopted conditions, the required effective $\alpha$ lies outside the physically motivated range near $z = 0$, preventing distance mimicry within the considered (m)ZKDR prescriptions. For ordinary ZKDR with the same fixed background parameters and \(w_0\ne-1\), reproducing the dynamical dark energy distance would require the smoothness parameter \(\alpha\) to become singular in the low-redshift limit, taking it outside the physically meaningful range \(0\leq\alpha\leq1\). For the modified prescription, the limit is finite but exceeds unity for the adopted CPL parameters. Thus, inhomogeneous light propagation in the presence of dynamical dark energy should be studied directly on a dynamical-dark-energy background rather than absorbed into a ZKDR smoothness parameter defined on a standard \(\Lambda\)CDM background.

\section*{Acknowledgments}
\noindent

The authors are grateful to S.V. Chervon and A.V. Nikolaev for valuable comments and discussions.

I.A. Moiseev is a recipient of a scholarship from the Basis Foundation for the Advancement of Theoretical Physics and Mathematics.

The research was supported by the Russian Science Foundation, project No. 25-22-20026.

\end{document}